\documentclass[twocolumn,showpacs,preprintnumbers,amsmath,amssymb]{revtex4}
\newcommand{\minus}[1]{ \mbox{$ (-1)^{#1} $}}

\newcommand{\ket}[1]{ \mbox{$|\, #1\,\rangle$}}

\newcommand{\ME}[3]{ \mbox{$\langle #1\,|\,#2\,|\,#3\rangle $} }

\newcommand{\MEred}[3]{ \mbox{$\langle #1\,||\,#2\,||\,#3\rangle $} }

\newcommand{\CGC}[6]{ \mbox{$ \left( #1 #2 , #3 #4\,|\, #5 #6 \right) $} }

\usepackage{dcolumn}
\usepackage{bm}
\usepackage{amsmath}
\usepackage{latexsym}
\usepackage{graphicx}
\usepackage{amsmath,amssymb,mathrsfs}
\usepackage{csquotes}
\usepackage{color}
\usepackage{amsbsy}

\begin{document}

\title{Photoelectron angular distribution in two-pathway ionization of neon with femtosecond XUV pulses}
\author{Nicolas Douguet$^1$, Elena~V.~Gryzlova$^2$, Ekaterina I.~Staroselskaya$^3$, Klaus Bartschat$^1$ Alexei~N.~Grum-Grzhimailo$^2$}
\affiliation{$^1$Department of Physics and Astronomy, Drake University, Des Moines, Iowa 50311, USA }
\affiliation{$^2$Skobeltsyn Institute of Nuclear Physics, Lomonosov Moscow State University, Moscow 119991, Russia}
\affiliation{$^3$Faculty of Physics, Lomonosov Moscow State University, Moscow 119991, Russia }

\begin{abstract} 
We analyze the photoelectron angular distribution in two-pathway interference between
non\-resonant one-photon and resonant two-photon ionization of neon. We consider a 
bichromatic femtosecond XUV pulse
whose fundamental frequency is tuned near the $2p^5 3s$ atomic states of neon.
The time-dependent Schr\"odinger equation is solved and the results are employed to compute the angular 
distribution and the associated anisotropy parameters at the main photoelectron line.
We also employ a time-dependent perturbative approach, which
allows obtaining information on the process for a large range of pulse parameters, 
including the steady-state case of continuous radiation, i.e., an infinitely long pulse.
The results from the two methods are in relatively good agreement over the domain of applicability 
of perturbation theory. 
\end{abstract}


\maketitle

\section{Introduction}
\label{intro}
Recent improvements in advanced light sources have open\-ed up a variety of new promising possibilities 
regarding the coherent control of atomic systems in the extreme ultra\-violet (XUV) energy range. 
One way to achieve coherent control of the photo\-electron angular distribution (PAD),
which has been experimentally realized with atoms 25 years ago by optical lasers, is to interfere one-photon
and two-photon ionization pathways~\cite{Baranova92,Yin92}. 
The two-photon and one-photon pathways are, respectively, produced by the fundamental and the second harmonic of the optical laser.
We will refer to this particular case as an $\omega + 2\,\omega$ process below.
This field originated in theory and was further developed experimentally \hbox{\cite{Baranova91,Schafer92,Wang01,Yamazaki07}}. 
Coherent control via two-pathway interference was reviewed, for instance, in~\cite{Ehlotzky01,Astapenko06}.

In certain cases, resonant ionization via an inter\-mediate state can be used to enhance the two-photon
pathway \hbox{\cite{Yin95,Grum15,Prince16,Douguet16}}.
The coherent XUV pulses from the Free-Electron Laser (FEL) at FERMI (Trieste, Italy) recently allowed
for the experimental manipulation of the PAD 
by controlling the relative time-delay between the fundamental and the second harmonic
of a linearly polarized  XUV femtosecond (fs) pulse to an unprecedented time resolution of 3.1~attoseconds (as)~\cite{Prince16}.
The experimental study employed neon as the atomic target, with one of the $(2p^5 4s)$ states with total 
electronic angular momentum $J=1$ as an inter\-mediate stepping stone,
by utilizing a two-color pulse of central wavelengths 63.0 and 31.5~nm, respectively.

In light of the experimental success and further expected investigations 
regarding coherent control of the PAD in neon, it is highly desirable
that such complex and expensive experiments are supported by theoretical efforts.
Therefore, one of the principal goals of the present work is to reveal
some of the main characteristics of the $\omega+2\,\omega$ process in neon,
this time choosing $(2p^5 3s)\,J=1$ as the inter\-mediate states. We picked the latter states for the present
study, since they can be reasonably well described in a non\-relativistic
$LS$-coupling scheme, while the two \hbox{$(2p^5 4s)\,J=1$} states require 
an inter\-mediate coupling description due to the lack of a
well-defined total spin.

In the theoretical work described below, we employed three different methods 
to describe the $\omega+2\,\omega$ process. In the first one, we solved
the time-dependent Schr\"{o}dinger equation (TDSE) numerically on a grid~\cite{Grum06},
using a single-active electron (SAE) potential to accurately represent the energies and
one-electron orbitals of neon. In the second method, we employed a time-dependent 
lowest (non\-vanishing) order perturbation theory (PT) approach with the target structure obtained from a 
multi-configuration Hartree-Fock (MCHF) calculation and only a few inter\-mediate states
accounted for in the second-order PT ionization amplitude.
Finally, we considered pulses with an infinite number of cycles~\hbox{(PT-$\infty$)} using a 
variationally stable method~\cite{Gao88,Orel88,Gao90,Staroselskaya15}, which effectively
accounts for all inter\-mediate states in the second-order PT ionization amplitude.

\begin{figure}
\resizebox{0.50\textwidth}{!}{%
\includegraphics[width=6cm,angle=-90]{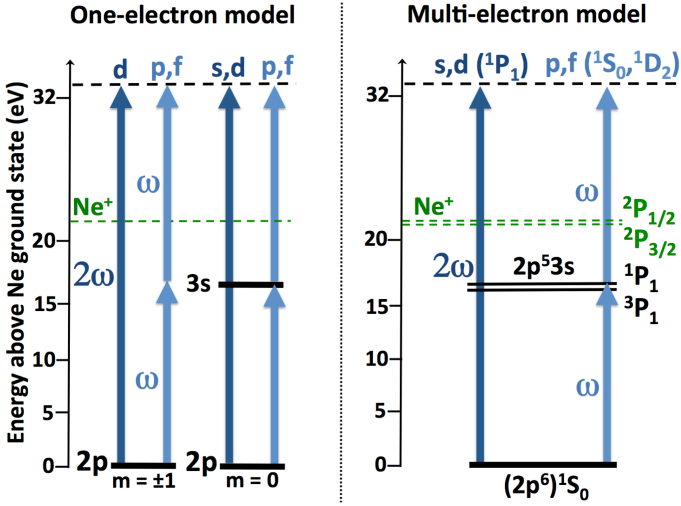}}
\caption{$\omega+2\,\omega$ ionization scheme by linearly polarized light in neon with $(2p^5 3s)\rm ^1P$ as the inter\-mediate state
in the single-active-electron model (left) and both $(2p^5 3s)\;J=1$ states in the multi-electron model (right).
See text for details.}
\label{fig:1}
\end{figure}

There exist two states, $2p^5{\rm (^2P_{3/2})}3s$ and
$2p^5{\rm (^2P_{1/2})}3s$, with total angular momentum $J=1$,
 which can be reached via optically allowed transitions from the $(2p^6)\rm ^1S_0$ initial state. 
As previously mentioned, these states are relatively well described in the $LS$-coupling scheme,
since they have predominant (93$\%$~\cite{Machado,NIST}) $\rm ^3P$ and $\rm ^1P$ character, respectively.
Therefore, we employ the $LS$-coupling scheme notations to label these states in the following development.


The $\omega+2\,\omega$ process using $(2p^5 3s)\,J=1$ as inter\-mediate states is presented in Fig.~\ref{fig:1}.
The scheme in the one-electron model is shown on the left panel, where we denote the electronic states
by listing only the active electron. Therefore, the inter\-mediate state (only the $\rm ^1P$ state is possible) 
is simply labelled $3s$, and this notation will be further used
throughout the manuscript. One-photon absorption of the second harmonic produces $s$- and $d$-wave photo\-electrons,
while two-photon absorption of the fundamental produces $p$- and $f$-wave photoelectrons. In the multi-electron
model (right panel of Fig.~\ref{fig:1}), these waves couple to the residual ionic state
to make the symmetries indicated at the top. The inter\-mediate $\rm ^3P_1$ and $\rm ^1P_1$ states,
corresponding to the $2p\to3s$ one-electron excitation, have, respectively, 16.67~eV and 16.85~eV excitation energies~\cite{NIST}. 
Since only $\rm ^1P_1$ can be efficiently excited, and it is well separated from other optically allowed states, 
it enables us to treat the effect of an ``almost'' isolated resonance. Consequently, it represents an excellent situation, 
with a minimum of additional complications, to compare results obtained by different models in a multi-electron system.

This manuscript is organized as follows.  In the next section,
we introduce our theoretical models, while Sect.~\ref{sec:2} is devoted to the presentation and analysis of our results.
Finally, Sect.~\ref{sec:3} contains our conclusions and perspectives for the future.
Unless indicated otherwise, atomic units are used throughout this manuscript.

\section{Theoretical approach}
\label{sec:1}
We consider a linearly polarized electric field of the form
\begin{equation}
E(t)=F(t)\left[\cos\omega t + \eta \cos (2\,\omega t+\phi)\right]\, ,
 \label{eq:field}
\end{equation}
where $\eta$ represents the amplitude ratio between the harmonics, $\phi$
is the carrier envelope phase (CEP) of the second harmonic, and $F(t)$ is
the envelope function. We employ the commonly used sine-squared envelope $F(t)=F_0\sin^2(\Omega t)$,
where $\Omega=\omega/2N$, with $N\gg1$ denoting the number of optical cycles. 

The details of our
TDSE approach can be found in~\cite{Grum15,Grum06,Abeln2010}.
The present TDSE calculations differ from our previous ones for electrons initially
in an $s$-orbital in that we now independently propagate the electronic wave packets
initially in the \hbox{$2p\;(m=0, \pm1)$} orbitals  and then average the results over the magnetic 
quantum numbers~$m$ to simulate an isotropic initial $2p^6\,\rm (^1S)$ state.
Here we only show briefly the main steps in the PT approach and describe
the physical models.

In second-order PT, the PAD for an initially unpolarized atom is given  by
\begin{equation}
\frac{d W}{d \Omega} = \frac{C}{2J_0+1} \sum_{M_0 \mu \atop{ J_f M_f}}
\left|  \eta \,U^{(1)}_{J_0M_0; J_f M_f, \vec{k} \mu} + U^{(2)}_{J_0M_0; J_f M_f,\vec{k} \mu} \right|^2 ,
\label{eq:ang_distr}
\end{equation}
where $\vec{k}$ is the linear momentum and $\mu$ the spin component 
of the photo\-electron, respectively; $J_0$ is the initial  total electronic angular momentum
with projection~$M_0$; $M_f$ is the projection of the residual ionic angular momentum~$J_f$;
$C$ is a normalization coefficient that is independent of the transition matrix elements and not relevant for our
further derivations. In Eq.~(\ref{eq:ang_distr}) we summed over~$J_f$, assuming incoherently
excited fine-structure levels of the residual ion. 

We choose the quantization $z$-axis along the electric field of the laser beams.
In the dipole approximation, the ionization amplitudes are given by
\begin{eqnarray}
U^{(1)}_{J_0M_0; J_f M_f,\vec{k} \mu} &=&-i \ME{J_f M_f, \vec{k} \mu^{(-)}}{D_z}{J_0M_0} \,T^{(1)} ,\label{eq:a1}\\
U^{(2)}_{J_0M_0; J_f M_f,\vec{k} \mu} &=& -\int \hskip-5.5truemm \sum_{~~n}
\ME{J_f M_f, \vec{k} \mu^{(-)}}{D_z}{\zeta_n J_n M_n} \nonumber\\
&&~~\times \ME{\zeta_n J_n M_n}{D_z}{J_0 M_0} \, T^{(2)}_{E_n}\label{eq:a2} \,.
\end{eqnarray}
Here $D_z = \sum_i d_{z,i} = \sum_i z_i $ is the $z$-component of the dipole operator, 
where the summation is taken
over all atomic electrons, and the sum (integral) in~(\ref{eq:a2}) is taken over all atomic
states with bound (continuum) energy $E_n$, labeled with their angular momentum~$J_n$, projection~$M_n$,
and the set of additional quantum numbers~$\zeta_n$. 
The values of the time integrals $T^{(1)}$ and $T^{(2)}_{E_n}$
were given in Eqs.~(9) and~(10) of Ref.~\cite{Grum15} (with the replacement $E_{1s} \to E_{2p}$).
The superscript~$(-)$ indicates the necessary asymptotic form of the continuum wave function, 
which is a distorted Coulomb wave calculated in the Hartree potential of the residual ion.

Upon expanding the ejected electron
wave function $\ket{\vec{k} \mu}^{(-)}$ in Eqs.~(\ref{eq:a1}) and~(\ref{eq:a2}) in (non\-relativistic)
partial waves and using standard angular momentum algebra, 
the PAD~(\ref{eq:ang_distr}) may be
written in the well-known form with Legendre polynomials $P_k(x)$ as
\begin{equation}
W(\theta) =  \frac{W_0}{4\pi}\left(1+\sum_k \beta_k P_k(\cos\theta)\right) \,,
\label{eq:ang_distr2} 
\end{equation}
with $W_0$ the angle-integrated cross section. The anisotropy parameters $\beta_k$ are generally given
by cumbersome expressions, including three types of terms, originating from
the first-order amplitude~(\ref{eq:a1}), the second-order amplitude~(\ref{eq:a2}),
and their interference.

In this paper, we are also interested in the differential asymmetry defined by~\cite{Grum15}
\begin{equation}
     A(0)  =  \frac{W(0)-W(\pi)}{W(0)+W(\pi)} =  \frac{\sum_{k=1,3,...} \beta_k}{1 + \sum_{k=2,4,...} \beta_k} \,.\label{eq:Asym} 
\end{equation}
As seen from the last part of the equation, a non\-zero asymmetry requires at least one non\-vanishing odd-rank anisotropy parameter.

In the non\-stationary PT version, 
we included seven inter\-mediate excited states of Ne in the sum over $n$ in Eq.~(\ref{eq:a2}), all with 
total angular momentum $J_n=1$: two states with configuration $2p^5 3s$, two states with $2p^5 4s$, and three
states with $2p^5 3d$. The final ionic states Ne$^+(2p^5)^2{\rm P}_{1/2,3/2}$ were treated
in the single-configuration Hartree-Fock approximation in the $LSJ$-coupling scheme.
The wave functions of the photo\-electron $\varepsilon \ell$ with energy $\varepsilon$ and orbital
angular momentum~$\ell$ were calculated in the Hartree-Fock frozen-core approximation~\cite{Froese97}.
For the states $\ket{\zeta_n J_n=1}$, we used the inter\-mediate-coupling scheme
and mixed the $2p^5 3s$, $2p^5 4s$, and $2p^5 3d$ configurations on the basis of the term-averaged atomic electron
orbitals.
For the PADs in a narrow range of photon energies around the excitation energy of the $2p^5 3s \, ^1{\rm P}_1$ state,
the effects of other $\ket{\zeta_n J_n=1}$ states, although being included,
are not expected to be very important, due to negligible admixtures
of other configurations and the weak violation of the $LS$-coupling. 

More compact expressions can be obtained for the $\beta_k$ parameters
in the single-configuration approximation and a pure $LS$-coupling scheme.
After transforming from multi-electron matrix elements to single-electron 
matrix elements \hbox{\cite{Sobelman72,Varshalovich88}}, we obtain for ionization from the closed
$2p^6$ shell of neon:
\begin{eqnarray} \label{eq:bk}
\beta_k = \beta^{(1)}_k + \beta^{(2)}_k + \beta^{(12)}_k \,, \qquad k =1, \,2, \,3, \,4.
\end{eqnarray}
The first term originates from the absolute square of
the first-order amplitude and contributes only for $k=2$:
\begin{eqnarray} \label{eq:bk1}
\beta^{(1)}_k & = & \frac{1}{N} \left| T^{(1)} \right|^2 \sum_{\ell \ell' m} 
\minus{m} \hat{\ell} \hat{\ell}' \CGC{\ell}{0}{\ell'}{0}{k}{0} \nonumber \\
 & & \times \CGC{\ell}{m}{\ell'}{-m}{k}{0} \CGC{1}{m}{\ell}{-m}{1}{0} 
 \CGC{1}{m}{\ell'}{-m}{1}{0}  \nonumber \\
 & & \times e^{i(\Delta_{\ell}^{(1)} - \Delta_{\ell'}^{(1)})} d_{\varepsilon\ell,2p} \,
 d_{\varepsilon\ell',2p}^{\ast} \,.
\end{eqnarray}
In accordance with the selection rules for angular momentum and 
parity, the summation in~(\ref{eq:bk1}) runs over 
$\ell =0, \,2$ and $\ell' = 0, \,2$ (at least one of $\ell$ or $\ell'$ should be non\-zero).
The second term in~(\ref{eq:bk}) originates from the absolute square of the second-order amplitude and contributes
for $k = 2, \,4$:
\begin{eqnarray} \label{eq:bk2}
\beta^{(2)}_k & = & \frac{\eta^2}{3N} \sum_{\ell \ell' m} 
\minus{m} \hat{\ell} \hat{\ell}' \CGC{\ell}{0}{\ell'}{0}{k}{0}  \CGC{\ell}{m}{\ell'}{-m}{k}{0}  \nonumber \\
 & &  \times e^{i(\Delta_{\ell}^{(2)} - \Delta_{\ell'}^{(2)})} \nonumber \\
 & & \times \;\int \hskip-7.5truemm \sum_{~~~n, \ell_n} \CGC{\ell_n}{m}{\ell}{-m}{1}{0} 
 \CGC{1}{m}{\ell_n}{-m}{1}{0} \nonumber \\ 
 & & ~~~~~~~~~~~~~~~~~~~~ \times T_{E_n}^{(2)} \,  d_{\varepsilon\ell,nl_n} \, d_{n\ell_n,2p} \nonumber \\
 & & \times \;\int \hskip-7.5truemm \sum_{~~~n'\!,\ell'_n} \CGC{\ell'_n}{m}{\ell'}{-m}{1}{0} 
 \CGC{1}{m}{\ell'_n}{-m}{1}{0} \nonumber \\ 
 & & ~~~~~~~~~~~~~~~~~~~~ \times T_{E_{n'}}^{(2) \ast} \,  d_{\varepsilon\ell',n' \ell'_n}^{\ast} \,
  d_{n'\ell_{n'},2p}^{\ast} \,.
 \end{eqnarray}
Here $\ell_n \!=\! 0, \, 2$; $\ell \!=\!1, \,3$; $\ell' \! =\! 1, \,3$.  
The third term in~(\ref{eq:bk}) represents the interference between the two amplitudes and contributes 
for $k \!=\!1, \, 3$:
\begin{eqnarray} \label{eq:bk3}
\beta^{(12)}_k & = & - \frac{2 \eta}{\sqrt{3} N} {\rm Re} \Big[ 
 \sum_{\ell \ell' m} \hat{\ell} \hat{\ell}' 
\CGC{\ell}{0}{\ell'}{0}{k}{0}  \CGC{\ell}{m}{\ell'}{-m}{k}{0}  \nonumber \\
& &  \times e^{i(\Delta_{\ell}^{(1)} - \Delta_{\ell'}^{(2)})} \CGC{1}{m}{\ell}{-m}{1}{0} 
T^{(1)} d_{\varepsilon\ell,2p} \nonumber\\
 & & \times \;\int \hskip-7.5truemm \sum_{~~~n, \ell_n} \minus{\ell_n} 
\CGC{\ell_n}{m}{\ell'}{-m}{1}{0} \CGC{1}{m}{\ell_n}{-m}{1}{0} \nonumber \\ 
 & & ~~~~~~~~~~~~~~~~~~~~ \times T_{E_n}^{(2) \ast} \,  d_{\varepsilon\ell',n \ell_n}^{\ast} \, d_{n\ell_n,2p}
\Big] \,,
\end{eqnarray}
with $\ell_n \!=\! 0, \, 2$; $\ell'_n \!=\! 0, \, 2$; $\ell \!=\!1, \,3$; $\ell' = 1, \,3$. 
Recall that non\-vanishing odd-rank anisotropy parameters such as $\beta_1$ and $\beta_3$ are responsible for a
non\-zero left-right asymmetry~(\ref{eq:Asym}).
Thus, non\-vanishing values of the asymmetry are due to interference between one-photon and
two-photon ionization amplitudes.

In Eqs.~(\ref{eq:bk1})-(\ref{eq:bk3})
\begin{eqnarray} \label{eq:n}
N & = & \left| T^{(1)} \right|^2 \sum_{\ell} \left|d_{\varepsilon\ell,2p} \right|^2 \nonumber \\ 
 &  & + \, \eta^2 \sum_{\ell m} \Big| \,\, \int \hskip-6.5truemm \sum_{~~n, \ell_n} \minus{\ell_n} \hat{\ell}^{-1}_n 
\CGC{\ell_n}{m}{\ell}{-m}{1}{0}  \nonumber \\
 &  &  ~~~~ \times \CGC{1}{m}{1}{0}{\ell_n}{m} \, T_{E_n}^{(2)} \, 
 d_{\varepsilon\ell,nl_n} \, d_{n\ell_n,2p} \Big|^2 \,.
\end{eqnarray}
Furthermore, ${\rm Re}[X]$ in Eq.~(\ref{eq:bk3}) denotes the real part of the complex quantity~$X$,
$\CGC{j_1}{m_1}{j_2}{m_2}{j_3}{m_3}$ is a Clebsch-Gordan coefficient, and $\hat{a} \equiv \sqrt{2a+1}$.
For convenience we separated the factor $i^{\ell} e^{i \delta_{\ell}}$ in the continuum wave function
($\delta_{\ell}$ is the scattering phase)
from the single-electron reduced dipole matrix elements \hbox{$d_{\varepsilon\ell,2p} = \MEred{\varepsilon\ell}{d}{2p}$}, 
$d_{\varepsilon\ell,nl_n}=\MEred{\varepsilon\ell}{d}{nl_n}$, \hbox{$d_{n\ell_n,2p} = \MEred{n \ell_n}{d}{2p}$},
and we introduced the abbreviations \hbox{$e^{i \Delta_\ell^{(1)}}\equiv i^{-\ell-1}e^{i \delta_\ell}$} and  
\hbox{$e^{i \Delta_\ell^{(2)}}\equiv -i^{-\ell} e^{i \delta_\ell}$}. For clarity, we
left the summations over the projections~$m$ in Eqs.~(\ref{eq:bk1})-(\ref{eq:n}) rather than working them out further
and expressing the final results in terms of \hbox{$nj$-symbols}.

Equations~(\ref{eq:bk})-(\ref{eq:n}) represent the SAE model within the PT for the
finite pulse duration. To turn to the limit of the infinite pulse duration 
($N\to\infty$, PT-$\infty$ model), the time factors $T^{(1)}$ and $T^{(2)}_{E_n}$ 
transform as~\cite{Grum15}
\begin{eqnarray} \label{eq:t1}
T^{(1)} & \to & \phantom{-} \frac{\sqrt{3} F_0}{4\sqrt{2}} \, e^{-i \phi} \,, \\
T^{(2)}_{E_n} & \to & - \frac{3 F^2_0}{32} \, \frac{i}{E_n - E_{2p} - \omega + i0}. \label{eq:t2}
\end{eqnarray}
These equations differ from Eqs.~(18) and~(19) of~\cite{Grum15} by an additional phase factor incorporated into the definition of~$\Delta$ 
and a factor~$\sqrt{3/8}$ that accounts for changing the total intensity from a finite $\sin^2$ pulse to a constant one of infinite length.

In the variationally stable method, which is applicable to an infinite pulse
duration, a variational procedure to find the extremum of a functional
is used instead of summing (integrating) over the infinite set of inter\-mediate states~$n$ 
in Eqs.~(\ref{eq:bk2})-(\ref{eq:n}) after substituting~(\ref{eq:t2}).
The method is described in detail in~\cite{Staroselskaya15,Gao89}.  
Varying the functional we expanded,
following~\cite{Gao88,Gao90,Staroselskaya15} and others,
the trial radial functions $\lambda(r)$ and $\mu(r)$
over the basis of Slater orbitals $\Phi_q(r) = N_qr^{\ell_n+q} e^{-\chi r}$ as
\begin{equation} \label{eq:lm}
\lambda(r) = \sum_{q=1}^Q a_q \Phi_q(r) \,, \quad 
\mu(r) = \sum_{q=1}^Q b_q \Phi_q(r) \,.
\end{equation}
Here $Q$ is the number of Slater's orbitals,
$N_q$ are normalization factors, the coefficients $a_q$ and $b_q$ are the
parameters to be varied, and $\chi$ is a constant whose value is taken to improve
convergence. In our case $Q=50$ and $\chi=2.5$. 
We used the electron wave functions of the initial and final states found in the 
Hartree-Fock-Slater~\cite{Herman63} 
local potential. The latter is also taken as the radial part of the atomic Hamiltonian 
when calculating the functional. 

Qualitatively, the behavior of the asymmetry parameters $\beta_k$ in the region
of the $3s$ state can be understood within a simplified PT approach, 
which includes only a single inter\-mediate $3s$ state. Using Eqs.~(\ref{eq:bk})-(\ref{eq:t2}),
the aniso\-tropy parameters $\beta_k$ may be reduced, for an infinite pulse, to simple parametric forms.
Similar formulas were
derived in~\cite{Grum15} for $\omega+2\,\omega$ ionization of an $s$-electron in the vicinity of
an isolated inter\-mediate state, where the active electron occupies a \hbox{$p$-orbital}.  Specifically:
\begin{eqnarray} \label{eq:fano}
\beta_1 & = &\frac{\epsilon}{\epsilon^2+1}B_1\cos(\phi+\phi_1)\, , \label{eq:beta1}\\
\beta_2 & = &B_2\frac{\epsilon^2}{\epsilon^2+1}+\frac{2}{\epsilon^2+1}\, , \label{eq:beta2}\\
\beta_3 & = &\frac{\epsilon}{\epsilon^2+1}B_3\cos(\phi+\phi_3)\, , \label{eq:beta3}\\
\epsilon & = &\frac{\Delta \omega}{\frac{1}{2}\Gamma_{\beta}}\, , \quad  \Delta \omega = \omega - (E_{3s}-E_{2p}) \,, \\
\Gamma_{\beta}& = &\frac{F_0|d_{\epsilon p,3s}d_{3s,2p}|}{2\eta\sqrt{2(|d_{\epsilon s,2p}|^2 + 
|d_{\epsilon d,2p}|^2)}}\, ,\label{eq:gamma}
\end{eqnarray}

\begin{eqnarray} \label{eq:fano2}
B_1 & = &\frac{\sqrt{3}|(2e^{i\Delta_s^{(1)}}d_{\epsilon s,2p}-\frac{4\sqrt{2}}{5}e^{i\Delta_d^{(1)}}d_{\epsilon d,2p})|}{\sqrt{|d_{\epsilon s,2p}|^2+|d_{\epsilon d,2p}|^2}}\, , \label{eq:b1} \\
B_2 & = &\frac{|d_{\epsilon d,2p}|^2-2\sqrt{2}\Re[e^{i(\Delta_s^{(1)} - \Delta_{d}^{(1)})}d_{\epsilon s,2p}d_{\epsilon d,2p}]}{|d_{\epsilon s,2p}|^2+|d_{\epsilon d,2p}|^2}\, , \label{eq:b2} \\
B_3 & = &\frac{6\sqrt{6}|d_{\epsilon d,2p}|}{5\sqrt{|d_{\epsilon s,2p}|^2+|d_{\epsilon d,2p}|^2}}\, , \label{eq:b3}\\
\phi_1& = &\arg\Bigg[\left(2e^{i\Delta_s^{(1)}}d_{\epsilon s,2p}-\frac{4\sqrt{2}}{5}
e^{i\Delta_d^{(1)}}d_{\epsilon d,2p}\right)\nonumber\\
 & &~~~~~~~~~\times e^{-i\Delta_p^{(2)}}d_{\epsilon p,3s}d_{3s,2p}\Bigg]\, , \label{eq:f1}\\
\phi_3 & = &\arg\left[-e^{i(\Delta_d^{(1)} - 
\Delta_{p}^{(2)})}d_{\epsilon d,2p}d_{\epsilon p,3s}d_{3s,2p}\right]\, . \label{eq:f3}
\end{eqnarray}
In the above simplified PT approach, $\beta_4 $, as well as all higher-rank 
anisotropy parameters, vanish.
It is somewhat surprising that the parameters $B_1,\, B_2,\, B_3$ depend neither on the 
two-photon amplitude nor on~$\eta$ in this simplified case. For infinite pulse duration within perturbation theory,
therefore, the maximum asymmetry and phase do not depend on the strength of the  second harmonic. 
However, the strength of the second harmonic affects the width of the asymmetry structure.
The above features will be clearly seen below in the results of particular PT-$\infty$ calculations.

\section{Results and discussion}
\label{sec:2}
The numerical calculations were performed for two sets of pulse parameters. In the first set,
we used a pulse~$\Pi_1$, with $N=250$ cycles and the second-harmonic intensity set to $1\%$ 
of the fundamental intensity, i.e., \hbox{$\eta=0.1$}.
In the second set, we used a longer pulse~$\Pi_2$, with $N=500$ cycles and a higher intensity ratio of the second harmonic
equal to 10$\%$ of the fundamental, i.e.,~$\eta=\sqrt{0.1}$ 
In both cases, the peak intensity of the fundamental was kept constant at 10$^{12}$W/cm$^2$.

\begin{figure}[t]
\resizebox{\columnwidth}{!}{%
  \includegraphics[width=1.5cm,angle=90]{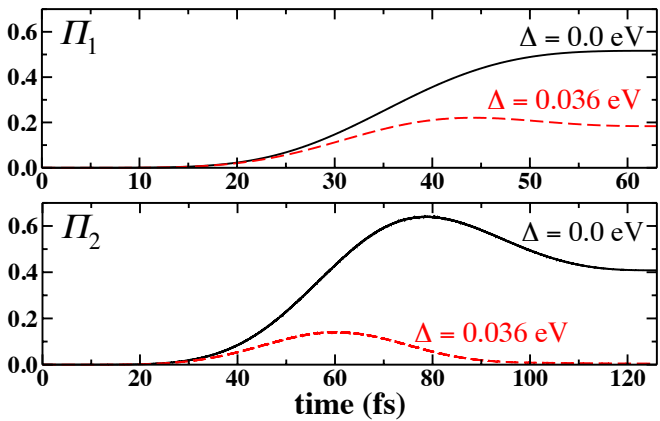}}
\caption{TDSE predictions for the population of the $3s$ state as a function of time. 
Results are shown for two different sets of pulse parameters.
The fundamental frequency~$\omega$ is either in resonance with the $3s$ inter\-mediate 
state \hbox{($\Delta=0.0$~eV)}, or slightly detuned
($\Delta=0.036$~eV).}
\label{fig:2}
\end{figure}

It is instructive to first analyze the efficiency of the femtosecond pulses in 
pumping the neon ground state to the excited $3s$ state.
Figure~\ref{fig:2} shows the evolution of the $3s$ state population as a function of time, 
for both~$\Pi_1$ and~$\Pi_2$, and for two different detunings~$\Delta$
of the fundamental frequency. 

The results behave in a predictable way. Focusing first on the resonant case \hbox{($\Delta=0.0$~eV)},
it is seen that the system does not even carry out half a Rabi oscillation for~$\Pi_1$, whereas, for the longer pulse~$\Pi_2$,
the system is close to have undergone one complete such oscillation. 
In both cases the $3s$ population can reach large values,
representing at their maximum, respectively, $51\%$ and $66\%$ of the total probability 
for~$\Pi_1$ and~$\Pi_2$. Note, however, that
a significant occupation of the $3s$ state occurs on rather different time scales for both pulses.
For example, the population of the $3s$ state is larger than $0.2$ for only $30$~fs during the pulse~$\Pi_1$, but
for more than $70$~fs during~$\Pi_2$. In the detuned case \hbox{($\Delta=0.036$~eV)}, the $3s$ population
decreases by slightly more than half for~$\Pi_1$ in comparison with the resonant case, 
whereas the population is drastically diminished for~$\Pi_2$.
This characteristic is readily understood: Since the spectral spread of~$\Pi_2$ is about half that of~$\Pi_1$,
the potential to drive an efficient population transfer decreases faster with increased detuning for~$\Pi_2$ than for~$\Pi_1$.

\begin{figure}[b]
\resizebox{\columnwidth}{!}{%
  \includegraphics[width=2.5cm,angle=90]{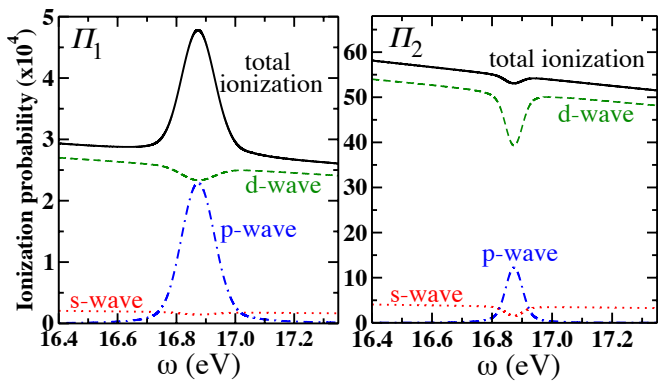}}
\caption{TDSE results for ionization to partial $s$-, $p$-, and \hbox{$d$-waves},
as well as total ionization at the main photoelectron line, for the~$\Pi_1$ and~$\Pi_2$ pulses.}
\label{fig:3}
\end{figure}

Turning now to the analysis of the ionization process, we present in Fig.~\ref{fig:3}
the partial-wave contributions for \hbox{$s$-}, $p$-, and $d$-waves, and the total ionization probability, 
at the main photoelectron line, for both pulses.
Out of resonance, both pulses exhibit similar characteristics, although ionization
is of course much more likely for the~$\Pi_2$ since (i)~it is a longer pulse
and (ii)~the strength of the second harmonic is ten times larger than for~$\Pi_1$.
In addition, the ionization out of resonance is strongly dominated by the $d$-wave,
representing more than $90\%$ of the total ionization probability in both cases.

On the other hand, one clearly observes two drastically different situations near resonance for each pulse.
In the first case ($\Pi_1$), while the second harmonic is weak at the resonance, 
the resonant $p$-wave ionization represents a large part of the (small) total ionization probability.
Note that the $p$-wave and $d$-wave contributions are nearly equal at resonance.
Consequently, a strong peak appears in the ionization spectrum when the fundamental frequency spans the resonance.
In the second case ($\Pi_2$), the second harmonic is so strong
that the background \hbox{$d$-wave} ionization strongly dominates the resonant
$p$-wave ionization. Since both $d$-wave and $s$-wave partial-wave 
ionization probabilities also decrease significantly at the resonance,
the total ionization probability barely reveals a fingerprint of the
resonance, apart from a small quenching.
The dip in the partial $s$-wave and $d$-wave ionization probabilities for~$\Pi_2$
at resonance is readily explained by the fact that the $2p$ orbital
is strongly depleted over time by efficient pumping from $2p$ to $3s$
via the fundamental frequency.

\begin{figure}[t]
  \includegraphics[width=3.8cm,angle=90]{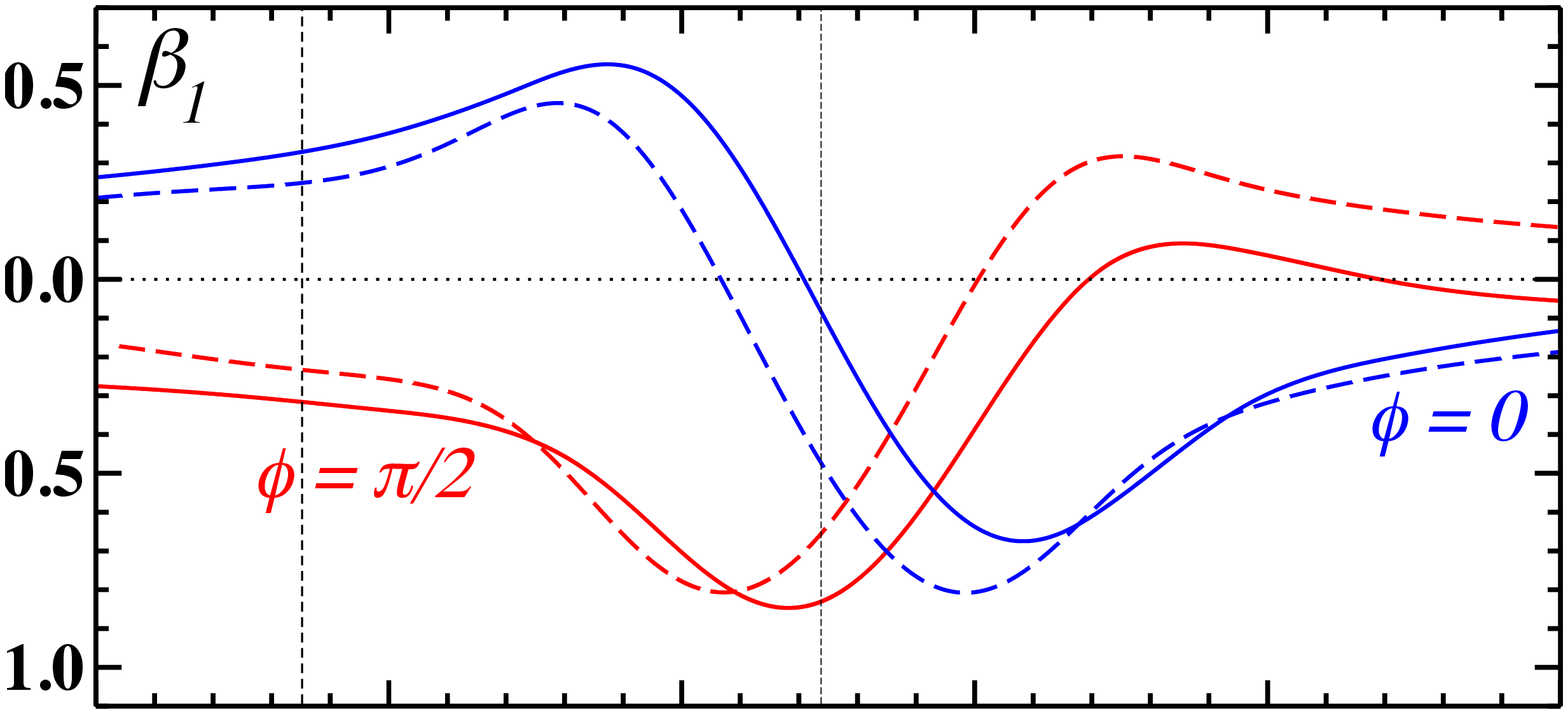}
  \includegraphics[width=3.78cm,angle=90]{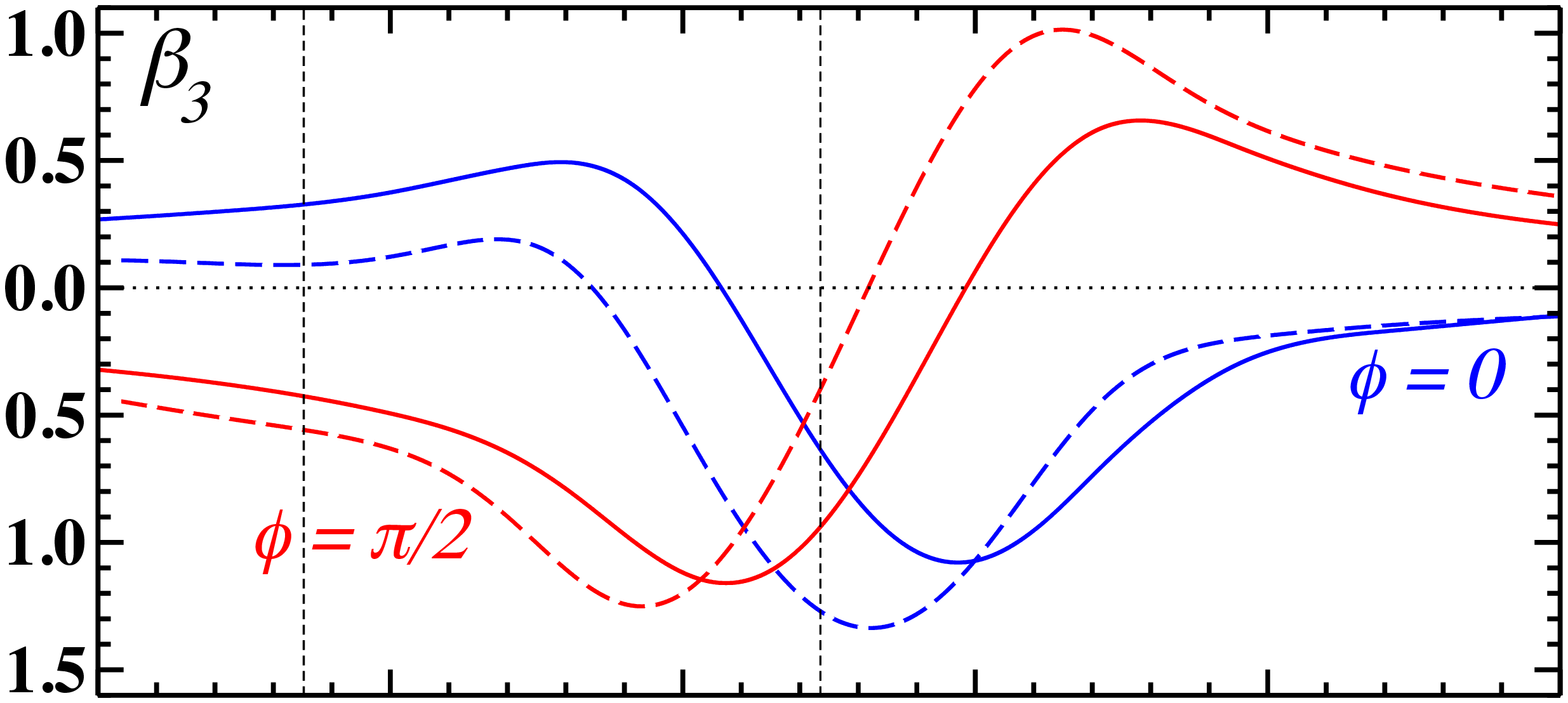}
  \includegraphics[width=4.55cm,angle=90]{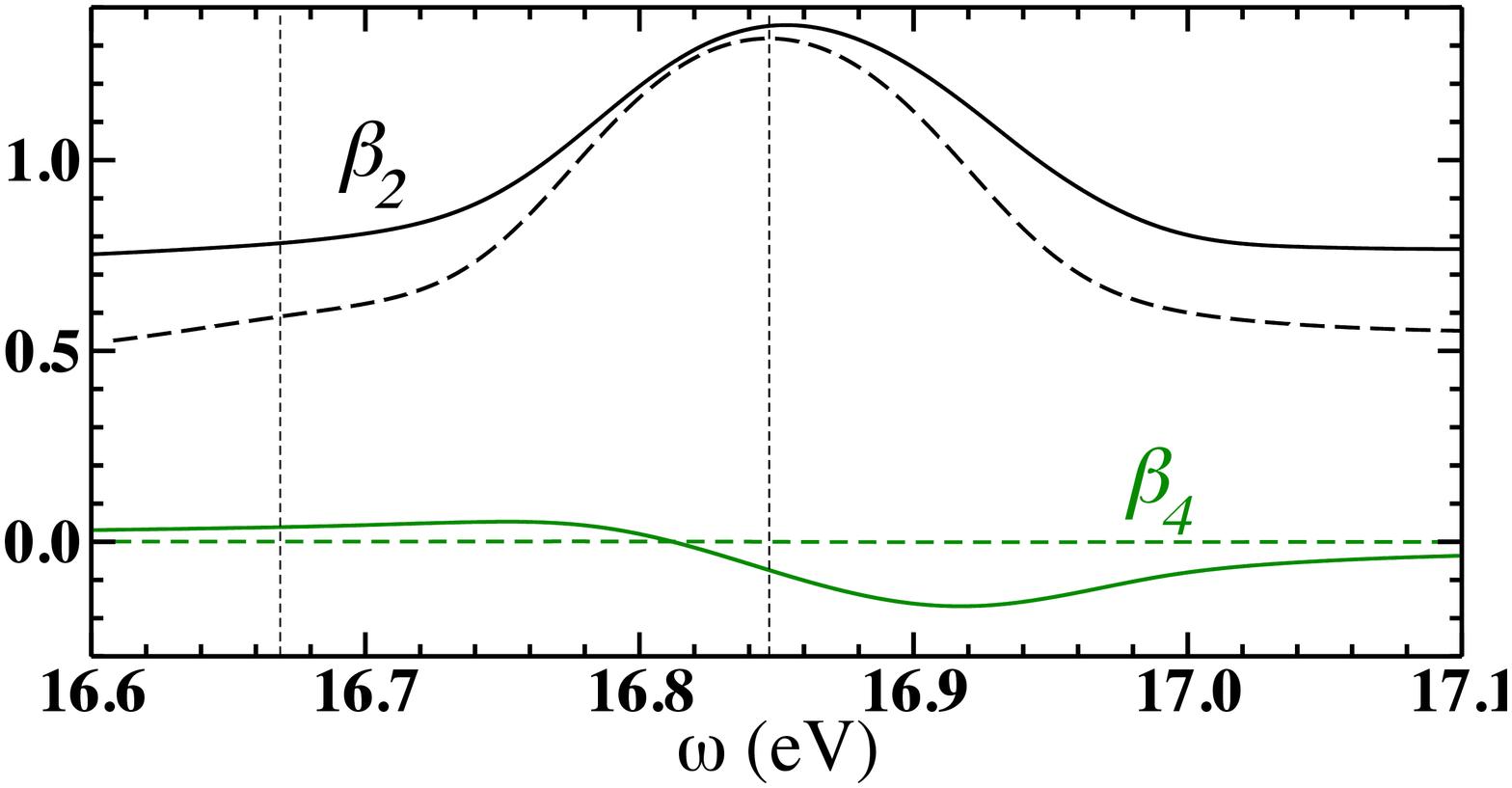}
  \caption{Anisotropy parameters~$\beta_1$,~$\beta_2$,~$\beta_3$, and~$\beta_4$, 
as a function of the fundamental frequency, for the pulse~$\Pi_1$.
Results are presented from both the TDSE (solid lines) and PT (dashed lines) approaches, 
and for CEPs $\phi=0$ and $\phi=\pi/2$ for~$\beta_1$ and~$\beta_3$.
The left and right vertical lines indicate the energy positions of the two $2p^5 3s$ states with $J=1$ 
and predominant triplet and singlet character, respectively.}
\label{fig:4}
\end{figure}

The calculated TDSE and PT values of the anisotropy parameters, introduced in Eq.~(\ref{eq:ang_distr2}), 
are presented in Fig.~\ref{fig:4} for ionization via~$\Pi_1$.
We only show~$\beta_k$  for $k \le 4$, since for the comparatively weak fields considered here,
 i.e., in the multi-photon regime, they represent the only significant non\-vanishing elements.
 We observe an overall satisfactory agreement between the TDSE and PT results.
 The odd-rank anisotropy parameters, presented for both cases of $\phi=0$ and $\phi=\pi/2$,
 exhibit an asymmetric Fano-like profile near resonance. Even far from resonance,~$\beta_1$ and~$\beta_3$
 assume non\-negligible values due to the spectral spread of~$\Pi_1$.
 The anisotropy parameter~$\beta_2$ peaks at resonance according
 to the increase in $p$-wave ionization (see Eq.~(\ref{eq:beta2})), while~$\beta_4$ becomes
 negligible everywhere. Note that the finite pulse duration leads not only to the broadening of
 the profile of the $\beta_1$ and $\beta_3$ parameters, but also to an energy shift of their zero crossing
 (see Eqs.~(\ref{eq:beta1}) and~(\ref{eq:beta3})) from the resonance position.

There are two principal reasons for the small discrepancies between the TDSE and PT
 results. To begin with, they can be attributed to the
 different electronic structure models in each approach. Recall that the 
 PT model uses a multi-electron MCHF description,
 while the solutions of the TDSE are obtained from a SAE potential. 
 On the other hand, it was shown that the population of 
 the $3s$ state can reach non\-negligible values,
 thus questioning the applicability of the PT approach. Despite these differences, the overall 
 satisfactory agreement obtained between the results from these two treatments
 suggests that the principal physics of the $\omega+2\,\omega$ process is properly accounted for in both approaches.
 
\begin{figure}[t]
  \includegraphics[width=3.8cm,angle=90]{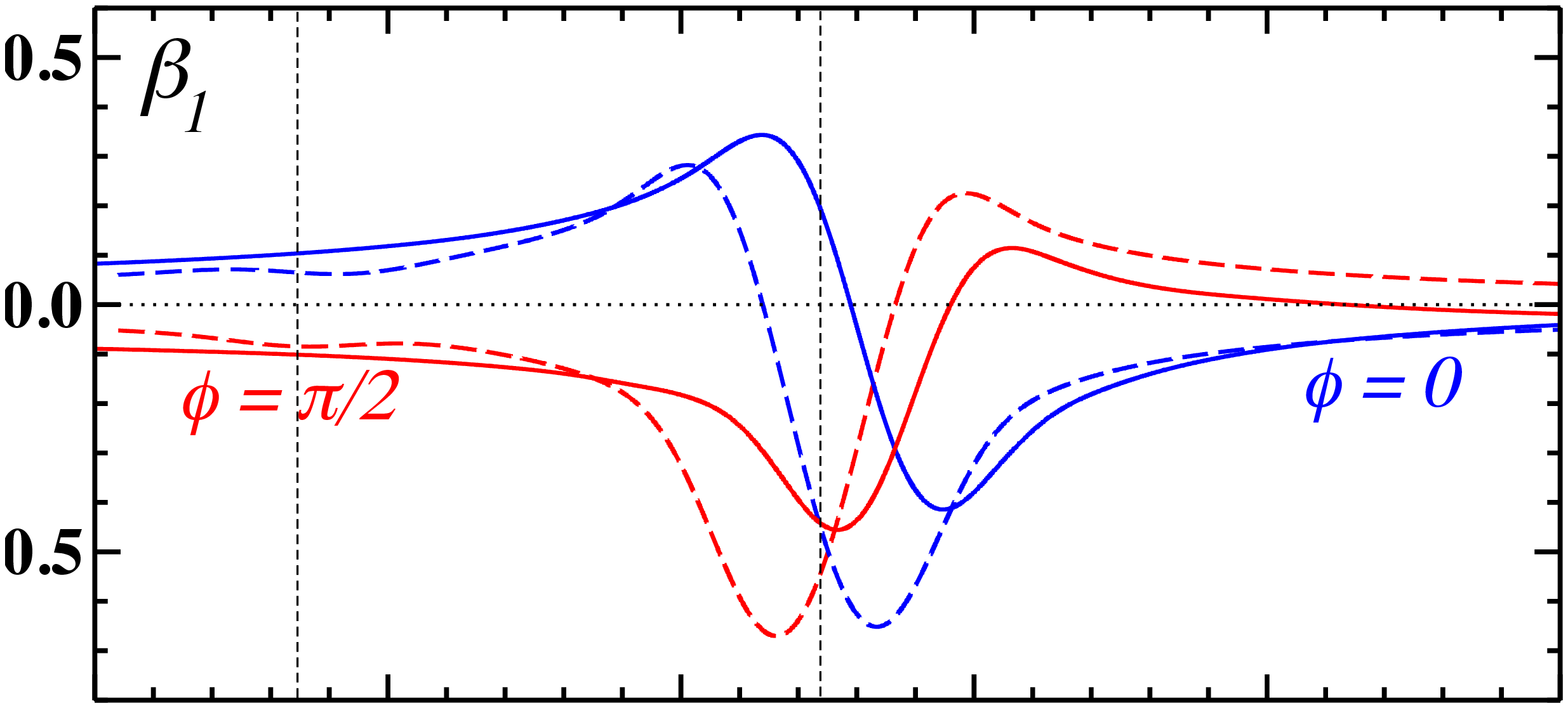}
  \includegraphics[width=3.8cm,angle=90]{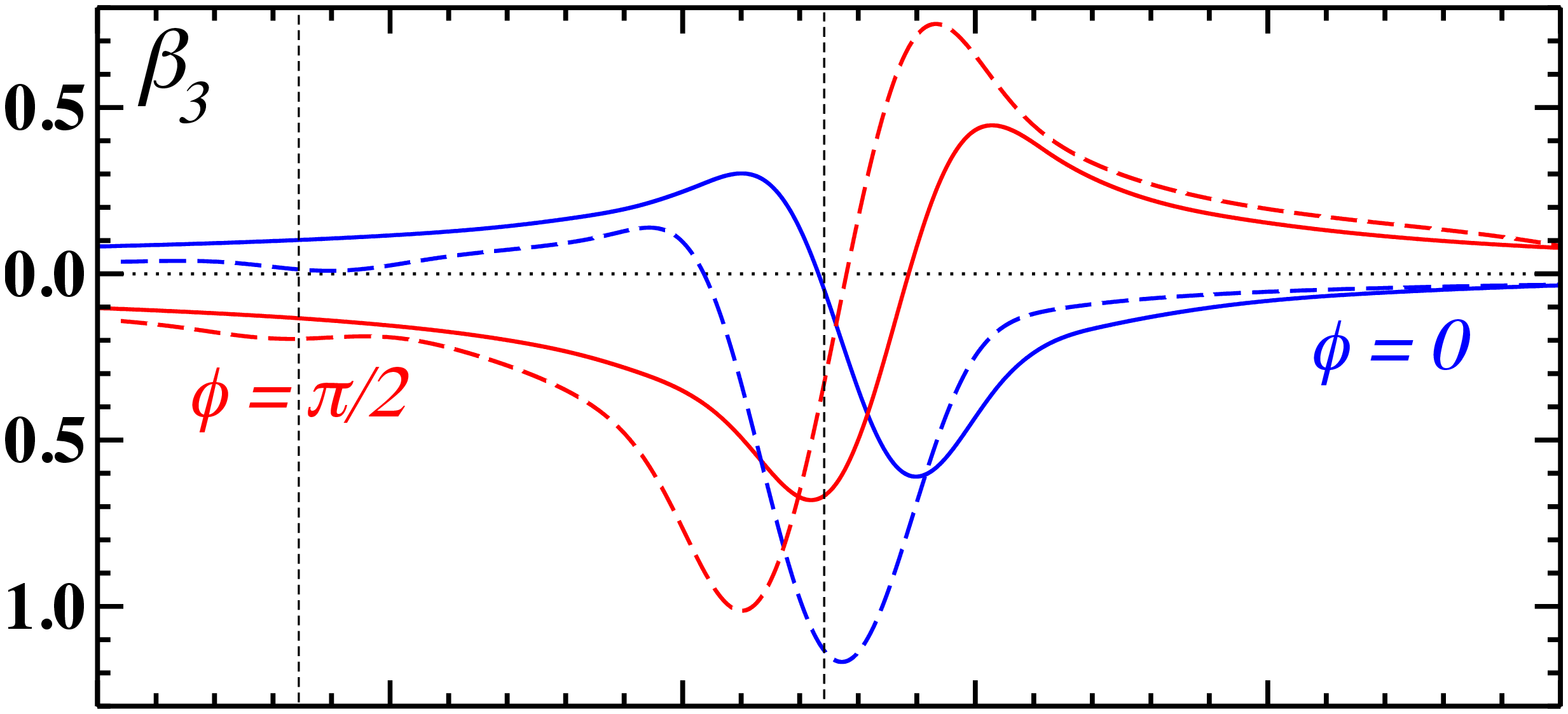}
  \includegraphics[width=4.53cm,angle=90]{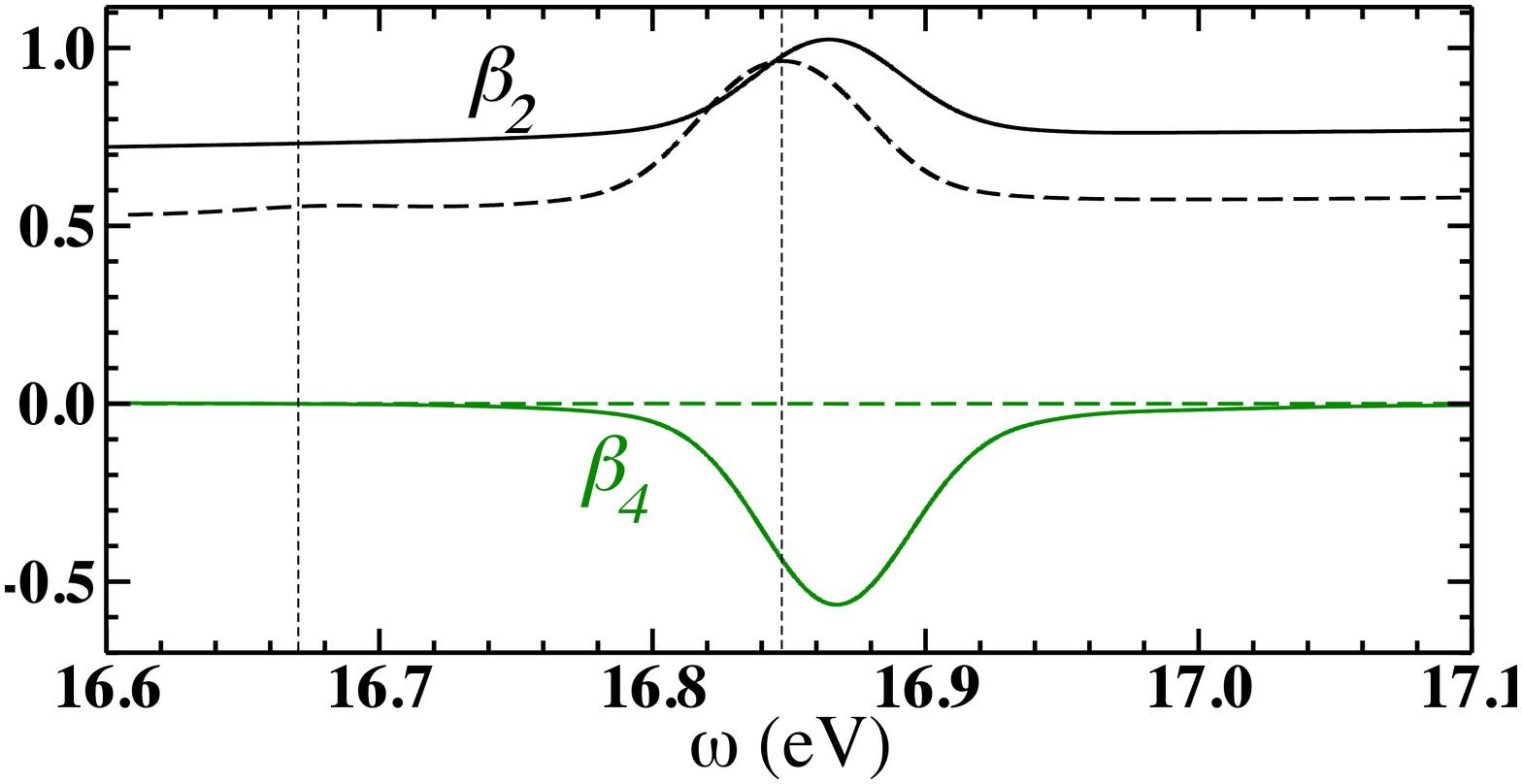}
\caption{Same as Fig.~\ref{fig:4} for the pulse~$\Pi_2$.}
\label{fig:5}
\end{figure}

The calculated anisotropy parameters associated with ionization by the pulse~$\Pi_2$ are 
presented in Fig.~\ref{fig:5}. Since the pulse is longer,
the width of the resonance profile is significantly narrower. As a result, 
the anisotropy parameters vary less and assume small values as
the fundamental frequency is detuned from resonance. The small effect of the 
other $3s$ state at $16.67$~eV can actually be noticed in the PT results.
Recall that this state is not included in the TDSE calculations. 
The values of~$\beta_1$ and~$\beta_3$ from TDSE and PT agree well
for detunings $\Delta\ge 0.1$~eV. On the other hand,
far from resonance, the~$\beta_2$'s calculated in each approach exhibit a small discrepancy from each other,
presumably due to the different structure models employed.
Here one can directly compare with experimental data to assess the validity of the results.
At $\omega=16.6$~eV, i.e., at $11.6$~eV photoelectron energy, the values of~$\beta_2$,
which are barely affected by resonance effects, are $0.72$, $0.53$, and $0.60$, respectively, 
for TDSE, PT, and PT-$\infty$.
At the same photoelectron energy a few groups~\cite{Codling76,Derenbach84,Schmidt86}
measured values of~$\beta_2$ in the interval approximately from 0.50 to 0.63, essentially
consistent with the predictions from all three models used in our study. The latter also agree
with Hartree-Fock~\cite{Kennedy72} and RPAE~\cite{Amusia72} calculations  
(see also the compilations in~\cite{Becker96,Schmidt97}).

\begin{figure}[t]
\begin{center}
\resizebox{\columnwidth}{!}{%
\includegraphics{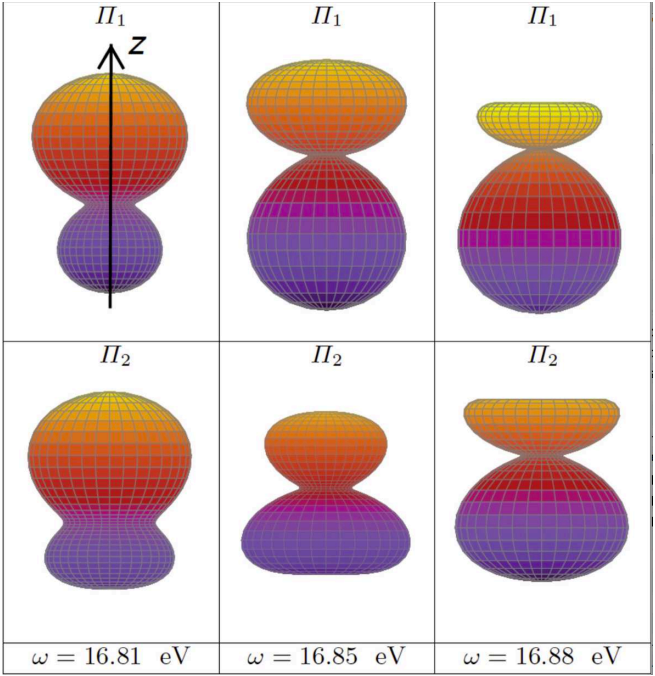}}
\caption{Calculated PADs in the TDSE approach for the pulse~$\Pi_1$ (top panels) and~$\Pi_2$ (bottom panels),
at three different fundamental frequencies~$\omega$ for $\phi=0$. The $z$-axis points upwards through the centers of the panels,
as indicated in the top left panel.
The distance from the center to a point on the surface is proportional to the probability density for the electron to be
ejected along this direction.}
\label{fig:6}
\end{center}
\end{figure}

Analyzing the results near resonance in Fig.~\ref{fig:5}, we observe that the TDSE and PT results
do not agree as well as for the shorter pulse~$\Pi_1$, even though the general trend looks similar. 
There is, however, an explanation for this
discrepancy. It can be understood by focusing first on the behavior of~$\beta_4$. 
There are two reasons for non\-vanishing values of~$\beta_4$. The first one is direct two-photon ionization into the $f$-wave channel.
However, $f$-wave ionization is almost negligible in all calculations presented in this work, and hence
interference of the $p$- and $f$- amplitudes only produces small non\-zero~$\beta_4$ values, which are visible in Fig.~\ref{fig:4} in the
TDSE calculation and also below in Fig.~\ref{fig:7}. 
A second, indirect reason for non\-zero~$\beta_4$ values in the present situation is the following: 
While the second harmonic ionizes neon, the fundamental frequency depletes
the $2p$ state over time, especially in the long $\Pi_2$ pulse. However, only the $m=0$ magnetic component of the $2p$ state 
can be pumped to the $3s$ state by the fundamental. As a consequence of the depletion of this sublevel, the 
second harmonic ionizes an \enquote{aligned} $2p$ state, thus leading to significant non\-vanishing 
values of~$\beta_4$. 
In our implementation of non\-stationary PT, $\beta_4\approx0$ as a result of cancellation of terms associated with different $d$-wave
components of a photoelectron emitted from an initially unpolarized target.
The second effect, therefore, cannot be 
accounted for in PT.  This explains most of the observed differences between the results from the two approaches.

It is also interesting to visualize the three-dimensional PAD for the two pulses considered.
\hbox{Figure~\ref{fig:6}} shows the PAD at the resonant frequency \hbox{($\omega=16.85$~eV)} and for small positive
($\omega=16.88$~eV) and negative ($\omega=16.81$~eV) detunings. For both pulses,
the direction of maximum emission along the electric field switches while passing through the resonance. 
The asymmetry of the PAD is relatively
large at the three considered frequencies for~$\Pi_1$. On the other hand, this asymmetry is less 
pronounced for~$\Pi_2$ due to the important 
contribution from~$\beta_4$, and the smaller contributions from~$\beta_1$ and~$\beta_3$.
The maximum electron emission forms an angle of about $145^\circ$ with the electric field 
at the resonance. 

 \begin{figure}
\resizebox{\columnwidth}{!}{%
\includegraphics{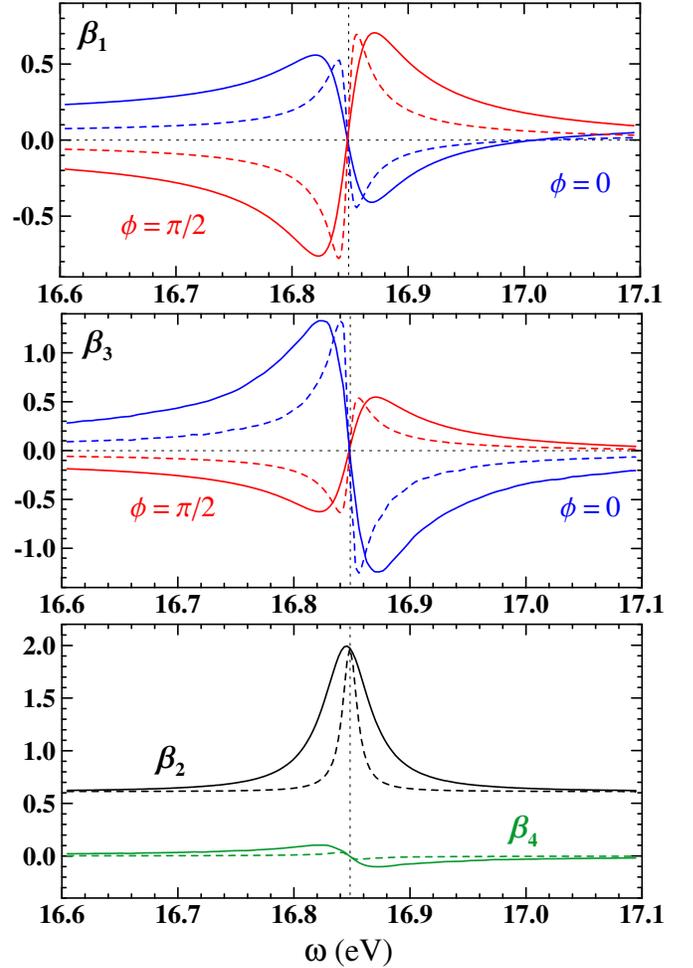}}
\caption{Same as Fig.~\ref{fig:4} for the PT-$\infty$ approach with infinite pulse duration and 
$\eta = 0.1$ (solid) and $\eta=\sqrt{0.1}$ (dashed).}
\label{fig:7}
\end{figure}

The results of the stationary PT-$\infty$ model are presented in Fig.~\ref{fig:7} for both amplitude 
ratios $\eta = 0.1$ and $\eta = \sqrt{0.1}$. 
The different anisotropy parameters vary according to the parametric forms given in Sect.~\ref{sec:1}. 
Accounting effectively for all intermediate states allows 
incorporating $f$-wave ionization more accurately in the PT-$\infty$ model than in non-stationary PT.
As a result, non\-zero, but still small $\beta_4$ values appear.

As predicted, the resonance profile is rather sharp for $N=\infty$ and the corresponding resonance 
would be broader for smaller~$\eta$, 
as seen in Eq.~(\ref{eq:gamma}). Other features predicted by Eqs.~(\ref{eq:beta1})--(\ref{eq:b3})
are very well seen: the odd-rank asymmetry parameters
go to zero at the resonance while~$\beta_2 \approx 2$ due to resonant excitation of  the inter\-mediate $3s$ state;
the amplitude of variations of $\beta_k$ ($k=1,\,2,\,3$) are independent of $\eta$.
Some of the strong variations of the anisotropy parameters are most likely exaggerated,
since in this specific case the population of the $3s$ state assumes significant values, 
thereby preventing an accurate description based on PT. 
Nevertheless, it is clear
that such a method has a predictive potential, particularly for weak fundamental intensities,
as it allows studying the currently realistic experimental situation of rather long FEL pulses.
Due to the computational resources required, such long pulses are particularly hard to simulate 
in the TDSE approach.

\begin{figure}
\resizebox{\columnwidth}{!}{%
\includegraphics{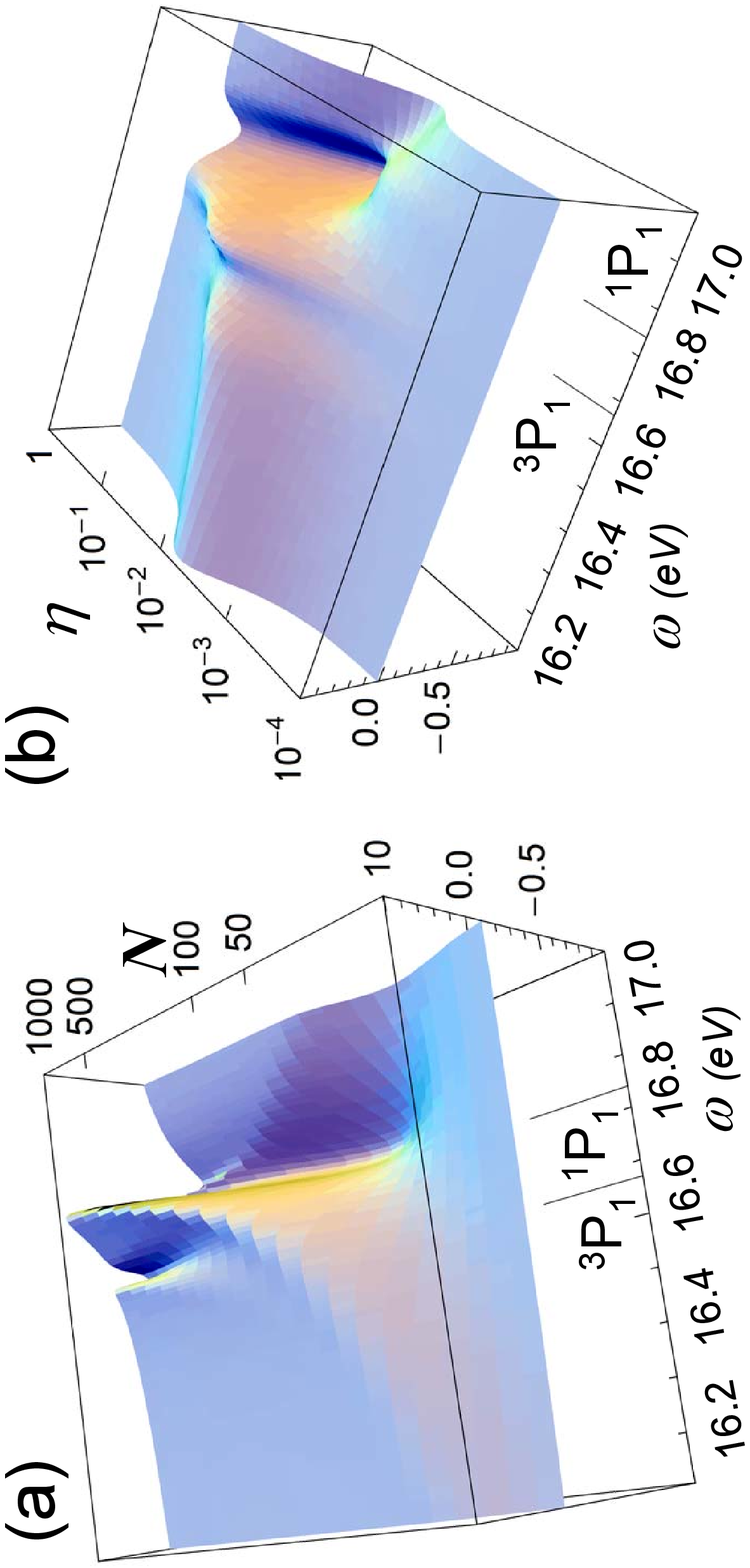}}
 \caption{Asymmetry $A(0)$ defined in Eq.~(\ref{eq:Asym}) for $\phi=0$ as function 
 of energy and number of cycles for $\eta=0.1$ and as function of energy and $\eta$ for $N=250$~(b).
 The positions of the two $2p^5 3s$ states with $J=1$ and predominant 
triplet and singlet character (see text) are indicated.}
\label{fig:8}
\end{figure}

An advantage of the PT approach is its ability to study 
the $\omega+2\,\omega$ process over a large range of pulse parameters with restricted
computational resources. 
Figure~\ref{fig:8} shows how the profile of the asymmetry~(\ref{eq:Asym})
is modified as a function of (a)~the number of cycles~$N$ 
and (b)~the relative strength of the second harmonic~$\eta$.
With increasing pulse duration the structures become sharper and more symmetric with respect to the $xy$-plane 
(maxima and minima are symmetric with respect to the resonance position and tend to be equal), 
and for pulse durations larger than $N=500$, the weak $2p^5(^2{\rm P}_{3/2})3s$ resonance becomes visible. 
This resonance (mainly of $\rm ^3P$ character) is excited very weakly, since it only contains 
a small ($\approx 7\%$) $\rm ^1P$ component. 
Nevertheless, when a long pulse is 
in resonance with $3s$, the interference with the second harmonic is noticeable. 

Figure~\ref{fig:8}b shows how the resonance profile changes with~$\eta$.
First, for increasing~$\eta$ the structure becomes narrower before the amplitude decreases. 
This is in contradiction with expectations from Eq.~(\ref{eq:gamma}) and Fig.~\ref{fig:7},
which predict a constant maximal amplitude.
However, for a finite pulse duration ($N=250$) and when $\Gamma_{\beta}$ is 
smaller than the pulse spectral width, this simple behavior breaks down.
For such a short pulse, there is only a small possibility to observe interference
with the weak $3s$ resonance: for small~$\eta$ the structure is very broad and one 
cannot distinguish the $(2p^5 3s)\rm ^1P_1$ resonance
from the tale of $(2p^5 3s)\rm ^3P_1$.  If~$\eta$ is too large, however, one-photon 
ionization dominates over the two-photon process, and there is practically no interference.

Finally, we note that within PT decreasing $\eta$ produces the same effect as increasing 
the total intensity.  This is due to the fact that the two-photon ionization amplitudes 
depend linearly on the intensity while the single-photon amplitude dependence is 
proportional to $\eta\,\sqrt{I}$. Although the PT simulation cannot be 
directly extrapolated to higher intensities, Fig.~\ref{fig:8} indicates that with
increasing intensity the profile of the asymmetry is broadening while its amplitude 
is decreasing.

\section{Conclusions and Future Perspectives}
\label{sec:3}
We analyzed in detail the ionization by a linearly polarized bichromatic XUV pulse containing the fundamental and the second harmonic 
($\omega+2\,\omega$ process) in neon 
using the \hbox{$(2p^5 3s)\;J=1$} states as inter\-mediate states. 
Two particular  situations were considered, corresponding to different pulse lengths and intensities 
of the second harmonic representing either $1\%$ or $10\%$ of the fundamental.
Solving the time-dependent Schr\"{o}dinger equation as well as employing a perturbative 
approach, we studied the variations of the photoelectron angular
distribution, in particular the aniso\-tropy para\-meters, as a function of the 
fundamental frequency. 

The results  demonstrate that for an inter\-mediate state of a multi-electron system, well separated 
from other electronic states and well represented in $LS$-coupling, theoretical models 
based on different approaches can achieve quantitative agreement in describing the characteristics of the
 $\omega+2\,\omega$ process. We also show that a significant asymmetry 
can be produced on a wide range of values of the amplitude ratio of the second harmonic over 
the fundamental. This is an interesting characteristic since the second harmonic admixture 
can sometimes be difficult to control, or even be estimated, experimentally.

Although results for both approaches agree well for the shorter pulse, noticeable effects
beyond the (lowest-order) perturbative approach can appear for longer pulse. In our case,
an unexpected resonant behavior of the $\beta_4$ parameter in the PAD was revealed
through the time-dependent calculations. Nevertheless, the perturbative treatment can
reproduce the main trends of the process and allows us to predict 
its outcomes for a variety of pulse parameters in the relatively weak-field regime,
especially for presently realistic experimental conditions, where relatively long pulses are employed.

Since the $\omega + 2\,\omega$ process using $(2p^5 4s)\,J=1$ as inter\-mediate states
was investigated experimentally, a thorough theoretical study of this case in the future
seems highly desirable. However, this situation presents additional challenges since the inter\-mediate states are not 
well described in the $LS$-coupling scheme.  Furthermore, high-lying discrete Rydberg as well as continuum states 
might play a significant role.

\section*{Acknowledgements}
The authors would like to acknowledge stimulating discussions with our experimental colleagues,
in particular Drs.\ G.~Sansone and K.~C.~Prince.
The work of ND and KB was supported by the United States National Science Foundation 
through grant No.\ \hbox{PHY-1403245}.  The TDSE calculations were performed on SuperMIC 
at Louisiana State University.  They were made possible by the XSEDE allocation
\hbox{PHY-090031}.

\end{document}